\documentclass[preprint, superscriptaddress,
showpacs,preprintnumbers,amsmath,amssymb,pra]{revtex4}
\usepackage{graphicx}
\usepackage{amsmath}\usepackage{slashed}

\begin{document}

\thispagestyle{empty}

\title{Casimir free energy of metallic films: Discriminating between Drude and plasma
model approaches}

\author{
G.~L.~Klimchitskaya}
\affiliation{Central Astronomical Observatory at Pulkovo of the Russian Academy of Sciences,
Saint Petersburg,
196140, Russia}
\affiliation{Institute of Physics, Nanotechnology and
Telecommunications, Peter the Great Saint Petersburg
Polytechnic University, St.Petersburg, 195251, Russia}

\author{
V.~M.~Mostepanenko}
\affiliation{Central Astronomical Observatory at Pulkovo of the Russian Academy of Sciences,
Saint Petersburg,
196140, Russia}
\affiliation{Institute of Physics, Nanotechnology and
Telecommunications, Peter the Great Saint Petersburg
Polytechnic University, St.Petersburg, 195251, Russia}

\begin{abstract}
We investigate the Casimir free energy of a metallic film either sandwiched between two
dielectric plates or in vacuum. It is shown that even for a thin film of several tens
of nanometer thickness the Casimir free energy and pressure calculated with the Lifshitz
theory using the Drude model and the plasma model approaches take significantly different
values and can be easily discriminated. According to our results, the classical limit
is already achieved for films of about 100\,nm thickness if the Drude model approach
is used in calculations. In this case the classical expressions for the Casimir free
energy and pressure are common for both configurations considered. If the plasma
model approach is used, the classical limit is not achieved for any film thickness.
Instead, the Casimir free energy and pressure are decreasing exponentially to zero.
When the plasma frequency goes to infinity, the Casimir free energy obtained using
the Drude model approach goes to a nonzero
limit in contradiction with expectations. If the plasma model approach is used the free
energy of metallic film goes to zero in the limit of infinitely large plasma frequency.
All analytic results are accompanied by numerical computations performed for a Au film
and sapphire plates. The possibilities to observe the predicted effects discriminating
between the Drude and plasma model approaches are discussed.
\end{abstract}

\maketitle

\section{Introduction}

During the last few years the Casimir effect, which manifests itself as free
energies and forces between closely spaced material boundaries, attracted much
experimental and theoretical attention \cite{1,2,3}.
The in-depth reason responsible for the Casimir effect is an existence of the
zero-point and thermal fluctuations of the electromagnetic field whose spectrum
is modified by the boundary conditions. Applications of the Casimir effect
extend from the nanoscale science \cite{4,5,6}, atomic physics \cite{7,8,9,10},
condensed matter physics \cite{11,12,13,14,15,16}, to the elementary particles,
astrophysics and cosmology \cite{17,18,19,20}.
The basic theory describing the Casimir effect is the Lifshitz theory of dispersion
forces \cite{3,21}. It was originally formulated for the plane parallel boundary
surfaces and recently generalized for the bodies of arbitrary geometrical
shape \cite{22,23}. This generalization was used for interpretation of experiments on
measuring the Casimir interaction between sinusoidally \cite{24,25,25a,25b} and
rectangular \cite{26,27} corrugated surfaces.

Calculations of the Casimir free energy and pressure using the Lifshitz theory
require the values of dielectric permittivities of boundary materials at the imaginary
Matsubara frequencies. The latter are obtained by means of the Kramers-Kronig relations
from the measured data for the frequency-dependent complex index of refraction.
Taking into account that these data are available only at frequencies exceeding some
minimum frequency $\omega_m$, they are usually extrapolated down to zero frequency
using some model \cite{3}. Theoretically, the most straightforward way of extrapolation
taking into account the relaxation properties of conduction electrons is by means
of the Drude model (the so-called Drude model approach). It was shown, however, that
the results of all precise experiments on measuring the Casimir interaction between
metallic surfaces, performed by means of micromachined oscillator \cite{28,29,30,31}
and atomic force microscope \cite{32,33,34,35}, exclude the predictions of the
Lifshitz theory using the Drude model approach at the confidence level up to 99.9\%.
The same measurement results were found to be consistent \cite{28,29,30,31,32,33,34,35}
with the predictions of the Lifshitz theory using the plasma model for extrapolation of
the optical data to zero frequency (i.e., the plasma model approach) which disregards
the relaxation properties of conduction electrons. Quantitatively, an agreement of the
measurement data with the plasma model approach at higher than 90\% confidence level was
demonstrated in Ref.~\cite{36}.

On the other hand, it was found that the Casimir entropy calculated for
metals with perfect crystal lattices
using the Drude model approach goes to a nonzero limit depending on the
parameters of a system when the temperature vanishes in violation of the third law of
thermodynamics, the Nernst heat theorem. This was proved for both nonmagnetic \cite{37,38,39}
and magnetic \cite{40} metals. The plasma model approach was shown to be in agreement with
the Nernst heat theorem \cite{37,38,39,40}. Thus, both the experimental data and
thermodynamics are surprisingly in favor of the model which should not be applicable at low,
quasistatic, frequencies and is usually used in the literature \cite{41} in the region
of infrared optic, where the relaxation processes do not play any role.
On the other hand, the Bohr-van~Leeuwen theorem, which states that the classical transverse
electromagnetic field has no influence on the matter in the state of thermal
equilibrium, was shown to be in agreement with the Drude model approach and in contradiction with
the plasma model approach \cite{42}. This conflict between the two theorems could
indicate that even in the classical limit, where the major contribution to the Casimir force
does not depend on the Planck constant, the quantum effects still remain important.
It should be also remembered that for dielectric test bodies the measured Casimir force of
several recent experiments agrees with theoretical predictions of the Lifshitz theory only if
the conductivity at a constant current (the dc conductivity) is omitted \cite{43,44,45,46,47}.
If the dc conductivity is included in calculation, the obtained
theoretical results are in contradiction
with the measurement data \cite{43,44,45,46,47} and violate the Nernst heat theorem
\cite{48,49,50,51}.

We emphasize that all precise experiments mentioned above were performed at short separation
distances below a micrometer between the test bodies. At these separations, differences
in theoretical predictions of the Drude and plasma model approaches  do not exceed
a few percent. In spite of the fact that the total measurement error was typically by an
order of magnitude lower, it is desirable to find the experimental configurations where the
differences in theoretical predictions of the two approaches were more sizable. In this regard,
an employment of large separation distances above $6\,\mu$m, where the predictions of both
approaches differ by a factor of two, is not helpful because the force magnitudes become too low.

To avoid this problem, Refs.~\cite{52,53,54} proposed the use of differential force measurements,
where theoretical predictions of the Drude model approach for the difference of two
forces are larger than those of the plasma model approach by up to a factor of
1000. The measurement results of this experiment have been reported recently \cite{55}.
They demonstrated an exclusion of the Drude model approach and consistency with the plasma
model one. Note, however, that for the plasma model approach the predicted magnitudes of the
force differences in this experiment are of the order of 0.1\,pN. To compare,
theoretical errors in the
calculated force differences are of the same order of magnitude.

In this paper we investigate the Casimir free energy and pressure for metallic films, either
in vacuum or sandwiched between two thick dielectric plates. We show that this configuration
possesses some unusual properties, as compared with the more standard geometries of two plates
interacting through a vacuum gap or a liquid intervening layer. Specifically, we demonstrate that
even for rather thin nonmagnetic metallic films (of several tens nanometer thickness) the
predictions of the Lifshitz theory using the Drude and plasma model approaches differ significantly
and can be easily discriminated.
Note that the Casimir energy of metallic films in vacuum was considered in
Refs.~[\onlinecite{57a,57b}]. However, this and below results were not obtained because all
computations using the Drude model have been performed only at zero temperature.
Next we show that with increasing film thickness the Casimir free
energy and pressure go to the classical limiting values which do not depend on the material
properties of metallic film and dielectric plates if the Drude model approach is used in calculations.
Unlike the standard geometries, the classical limit is already achieved for the film of about
110\,nm thickness. For the plasma model approach, the dependence on the material properties of the
film and the plates is preserved up to relatively large film thicknesses. In this case, with
increasing film thickness, the Casimir free energy and pressure are
decreasing exponentially to zero and the differences with theoretical
predictions of the Drude model approach reach several orders of magnitude.
It is shown that the Casimir free energy of a metallic film described by the Drude model
approach  goes to a nonzero classical value when the plasma frequency goes to infinity.
This is contrary to physical intuition because the field fluctuations cannot penetrate inside an
ideal metal. If the plasma model approach is used, the free energy of a metallic film goes to
zero  when the plasma frequency goes to infinity. All the analytic results are illustrated by
numerical computations performed for the example of a Au film
situated either in vacuum or sandwiched
between two Al${}_2$O${}_3$ (sapphire) plates.

The paper is organized as follows. In Sec.~II we briefly formulate the basic equations of the
Lifshitz theory adapted for our configuration and present the analytic results for the Casimir
free energy. Section~III is devoted to numerical computations of the Casimir free energy of
a Au film in vacuum or between two sapphire plates of different thicknesses and temperatures.
In Sec.~IV similar results are obtained for the Casimir pressure. Section~V contains our
conclusions and discussion of the Drude and plasma model approaches. In Appendix some details
of our analytic evaluations are presented.

\section{Casimir free energy of metallic films}

We consider the three-layer system consisting of a thick plate (semispace),  described by the
dielectric permittivity $\varepsilon^{(-1)}(\omega)$, followed by a metallic film of thickness
$a$ described by the dielectric permittivity $\varepsilon^{(0)}(\omega)$ and another
thick plate (semispace) characterized by the dielectric permittivity $\varepsilon^{(+1)}(\omega)$.
The system is assumed to be in thermal equilibrium at temperature $T$. The Lifshitz formula for
the Casimir free energy per unit area is given by \cite{3,21}
\begin{eqnarray}
&&
{\cal F}(a,T)=
\frac{k_BT}{2\pi}\,\sum_{l=0}^{\infty}{\vphantom{\sum}}^{\prime}
\int_{0}^{\infty}\!\!\!k_{\bot}\,dk_{\bot}
\left\{\ln\left[1-r_{\rm TM}^{(0.+1)}(i\xi_l,k_{\bot})
r_{\rm TM}^{(0.-1)}(i\xi_l,k_{\bot})
e^{-2ak_l^{(0)}(k_\bot)}\right]\right.
\nonumber \\
&&~~~~~~~~~~~
+\left.\ln\left[1-r_{\rm TE}^{(0.+1)}(i\xi_l,k_{\bot})
r_{\rm TE}^{(0.-1)}(i\xi_l,k_{\bot})
e^{-2ak_l^{(0)}(k_\bot)}
\right]\right\},
\label{eq1} 
\end{eqnarray}
\noindent
where  $k_B$ is the Boltzmann constant,
$\xi_l=2\pi k_BTl/\hbar$ with $l=0,\,1,\,2,\,\ldots$ are the
Matsubara frequencies,
$k_{\bot}=|\mbox{\boldmath$k$}_{\bot}|$ is the magnitude of the
projection of the wave vector on the plane of plates,
and the prime  multiples the term with $l=0$ by 1/2.
The reflection coefficients  for two independent
polarizations of the electromagnetic field, transverse magnetic (TM) and
transverse electric (TE), are expressed as
\begin{eqnarray}
&&
r_{\rm TM}^{(0,\pm 1)}(i\xi_l,k_{\bot})=\frac{\varepsilon_{l}^{(\pm 1)}
k_{l}^{(0)}(k_{\bot})-\varepsilon_{l}^{(0)}
k_l^{(\pm 1)}(k_{\bot})}{\varepsilon_{l}^{(\pm 1)}
k_{l}^{(0)}(k_{\bot})+\varepsilon_{l}^{(0)}
k_l^{(\pm 1)}(k_{\bot})},
\nonumber \\
&&
r_{\rm TE}^{(0,\pm 1)}(i\xi_l,k_{\bot})=\frac{k_{l}^{(0)}(k_{\bot})-
k_l^{(\pm 1)}(k_{\bot})}{k_{l}^{(0)}(k_{\bot})+
k_l^{(\pm 1)}(k_{\bot})},
\label{eq2}
\end{eqnarray}
\noindent
where the following notation is introduced:
\begin{equation}
k_l^{(n)}(k_{\bot})=\sqrt{k_{\bot}^2+
{\varepsilon_{l}^{(n)}}\frac{\xi_l^2}{c^2}}
\label{eq3}
\end{equation}
\noindent
and $\varepsilon_{l}^{(n)}\equiv\varepsilon^{(n)}(i\xi_l)$ with $n=0,\,\pm 1$.

If both thick plates are made of common material, one should but
$\varepsilon_l^{(-1)}=\varepsilon_l^{(+1)}$ in Eqs.~(\ref{eq2}) and (\ref{eq3}).
In this case we have
$r_{\rm TM,TE}^{(0,+1)}(i\xi_l,k_{\bot})=r_{\rm TM,TE}^{(0,-1)}(i\xi_l,k_{\bot})$.
Below we also consider the Casimir free energy for a metallic film in vacuum.
This case is obtained from Eqs.~(\ref{eq2}) and (\ref{eq3}) by putting
$\varepsilon_{l}^{(-1)}=\varepsilon_{l}^{(+1)}=1$.
As a result, Eq.~(\ref{eq2}) takes the form
\begin{eqnarray}
&&
r_{\rm TM}^{(0,+1)}(i\xi_l,k_{\bot})=
r_{\rm TM}^{(0,-1)}(i\xi_l,k_{\bot})=\frac{k_{l}^{(0)}(k_{\bot})
-\varepsilon_{l}^{(0)}
q_l(k_{\bot})}{k_{l}^{(0)}(k_{\bot})+\varepsilon_{l}^{(0)}q_l(k_{\bot})},
\nonumber \\
&&
r_{\rm TE}^{(0,+1)}(i\xi_l,k_{\bot})=
r_{\rm TE}^{(0,-1)}(i\xi_l,k_{\bot})=\frac{k_{l}^{(0)}(k_{\bot})-
q_l(k_{\bot})}{k_{l}^{(0)}(k_{\bot})+q_l(k_{\bot})},
\label{eq4}
\end{eqnarray}
\noindent
where
\begin{equation}
q_l(k_{\bot})=\sqrt{k_{\bot}^2+\frac{\xi_l^2}{c^2}}.
\label{eq5}
\end{equation}

The analytic results of this section are obtained for metallic films described by the
Drude or by the plasma model (the interband transitions of core electrons are taken into
account in numerical computations performed in Secs.~III and IV).
Thus, the dielectric permittivity of the Drude model at the imaginary Matsubara frequencies
 is given by
\begin{equation}
\varepsilon_{l,D}^{(0)}=1+
\frac{{\omega}_p^2}{\xi_l[\xi_l+{\gamma}(T)]},
\label{eq6}
\end{equation}
\noindent
where $\omega_p$ is the plasma frequency and ${\gamma}(T) $ is the relaxation parameter.
The plasma model is obtained from Eq.~(\ref{eq6}) by putting the relaxation parameter
equal to zero
\begin{equation}
\varepsilon_{l,p}^{(0)}=1+
\frac{{\omega}_p^2}{\xi_l^2}.
\label{eq7}
\end{equation}

As mentioned in Sec.~I, the Drude model (\ref{eq6}) takes into account the relaxation
properties of conduction electrons. In classical electromagnetic fields it is applicable
at low frequencies. The plasma model (\ref{eq7}) disregards the relaxation
properties of conduction electrons. In classical fields it is applicable in the region
of infrared optics. Both quantities (\ref{eq6}) and (\ref{eq7}) can be continued to the
plane of complex frequencies as the analytic functions satisfying all the demands required
from the dielectric permittivity \cite{41}. Specifically, they satisfy the Kramers-Kronig
relations between their real and imaginary parts formulated for the functions having the first-
and second-order poles at zero frequency, respectively \cite{56}.

We begin from calculation of the zero-frequency contribution to Eq.~(\ref{eq1}) in the
case of the Drude model
\begin{equation}
{\cal F}_{D}^{(l=0)}(a,T)={\cal F}_{D,\,\rm TM}^{(l=0)}(a,T)+
{\cal F}_{D,\,\rm TE}^{(l=0)}(a,T).
\label{eq8}
\end{equation}
\noindent
Using Eq.~(\ref{eq6}), we obtain from Eq.~(\ref{eq3}) that
$k_0^{(n)}(k_{\bot})=k_{\bot}$ and Eq.~(\ref{eq2}) results in
\begin{equation}
r_{\rm TM}^{(0,\pm 1)}(0.k_{\bot})=-1, \qquad
r_{\rm TE}^{(0,\pm 1)}(0.k_{\bot})=0.
\label{eq9}
\end{equation}
\noindent
Then, using Eqs.~(\ref{eq1}) and (\ref{eq8}), we obtain
\begin{eqnarray}
&&
{\cal F}_{D,\,\rm TM}^{(l=0)}(a,T)=\frac{k_BT}{4\pi}\int_{0}^{\infty}\!\!\!
k_{\bot}dk_{\bot}\ln\left(1-e^{-2ak_{\bot}}\right)=
-\frac{k_BT}{16\pi a^2}\zeta(3),
\nonumber \\
&& 
{\cal F}_{D,\,\rm TE}^{(l=0)}(a,T)=0,
\label{eq10}
\end{eqnarray}
\noindent
where $\zeta(z)$ is the Riemann zeta function.
This result does not depend on the type of dielectric materials of the plates on both
sides of the metallic film and metal used. Note that the same result (the so-called
classical limit) is obtained for the well studied configuration of two metallic
semispaces separated by a dielectric film \cite{3}. As shown below, however, in the
present case of metallic film the classical limit is achieved at much shorter
separations than for two metallic semispaces.

Now we consider the case of plasma model (\ref{eq7}). Here, like for the Drude model,
$k_0^{(\pm 1)}(k_{\bot})=k_{\bot}$ but
\begin{equation}
k_0^{(0)}(k_{\bot})=\sqrt{k_{\bot}^2+\frac{\omega_p^2}{c^2}}.
\label{eq11}
\end{equation}
\noindent
As a result,
\begin{equation}
r_{\rm TM}^{(0,\pm 1)}(0.k_{\bot})=-1, \qquad
r_{\rm TE}^{(0,\pm 1)}(0.k_{\bot})=
\frac{\sqrt{k_{\bot}^2+\frac{\omega_p^2}{c^2}}-
k_{\bot}}{\sqrt{k_{\bot}^2+\frac{\omega_p^2}{c^2}}+k_{\bot}}
\label{eq12}
\end{equation}
\noindent
and for the TM contribution to the Casimir free energy defined like in
Eq.~(\ref{eq8}) one obtains
\begin{equation}
{\cal F}_{p,\,\rm TM}^{(l=0)}(a,T)=\frac{k_BT}{4\pi}\int_{0}^{\infty}\!\!\!
k_{\bot}dk_{\bot}\ln\left(1-e^{-2a\sqrt{k_{\bot}^2+\frac{\omega_p^2}{c^2}}}\right)
=-\frac{k_BT}{4\pi}\sum_{n=1}^{\infty}\frac{1}{n}\int_{0}^{\infty}\!\!\!
k_{\bot}dk_{\bot}e^{-2an\sqrt{k_{\bot}^2+\frac{\omega_p^2}{c^2}}}.
\label{eq13}
\end{equation}
\noindent
Calculating the integral in Eq.~(\ref{eq13}), we finally have
\begin{equation}
{\cal F}_{p,\,\rm TM}^{(l=0)}(a,T)=-\frac{k_BT}{16\pi a^2}
\sum_{n=1}^{\infty}\frac{1}{n^3}\left(1+\frac{2an\omega_p}{c}\right)
e^{-2an{\omega_p}/{c}}
=-\frac{k_BT}{16\pi a^2}\left[\tilde{\omega}_p\,{\rm Li}_2(e^{-\tilde{\omega}_p})
+{\rm Li}_3(e^{-\tilde{\omega}_p})\right],
\label{eq14}
\end{equation}
\noindent
where ${\rm Li}_n(z)$ is the polylogarithm  function and
$\tilde{\omega}_p=2a\omega_p/c$.
As is seen in Eq.~(\ref{eq14}), in our configuration the TM contribution to
${\cal F}_{p}^{(l=0)}$ is not similar to Eq.~(\ref{eq10}) and decreases exponentially
with increasing film thickness. The physical meaning of this dependence is
explained below.

The TE contribution to the zero-frequency term in the case of the plasma model is a more
complicated quantity. To calculate it, we use Eqs.~(\ref{eq1}), (\ref{eq11}),
(\ref{eq12}) and introduce the new dimensionless variable $u=2ak_{\bot}$
with a result
 \begin{equation}
{\cal F}_{p,\,\rm TE}^{(l=0)}(a,T)=\frac{k_BT}{16\pi a^2}
\int_{0}^{\infty}\!\!\!udu\ln\left[1-r_{\rm TE}^2(0,u)
e^{-\sqrt{u^2+\tilde{\omega}_p^2}}\right],
\label{eq15}
\end{equation}
\noindent
where
\begin{equation}
r_{\rm TE}(0,u)\equiv r_{\rm TE}^{(0,\pm 1)}(0,u)=
\frac{\sqrt{u^2+\tilde{\omega}_p^2}-u}{\sqrt{u^2+\tilde{\omega}_p^2}+u}.
\label{eq16}
\end{equation}

Now we notice that $1/\tilde{\omega}_p=c/(2a\omega_p)=\delta/(2a)$. Here,
$\delta=c/\omega_p=\lambda_p/(2\pi)$, where $\lambda_p$ is the plasma wavelength,
has the meaning of the effective penetration depth of the electromagnetic oscillations
into a metal. For Au, for instance, $\omega_p=9\,$eV \cite{57} and
$\delta\approx 22\,$nm. Thus, at $a\geq 110\,$nm the quantity $1/\tilde{\omega}_p$
can be considered as a small parameter. Expanding the pre-exponent in Eq.~(\ref{eq15})
in powers of this parameter, one obtains
\begin{equation}
r_{\rm TE}^2(0,u)=1-
\frac{4u}{\tilde{\omega_p}}+\frac{8u^2}{\tilde{\omega}_p^2}.
\label{eq17}
\end{equation}
\noindent
Taking into account that for the same film thickness $\exp(-\sqrt{u^2+\tilde{\omega}_p^2})$
is even much smaller parameter, one can expand the logarithm in Eq.~(\ref{eq15}) and
find using Eq.~(\ref{eq17})
\begin{equation}
\ln\left[1-r_{\rm TE}^2(0,u)
e^{-\sqrt{u^2+\tilde{\omega}_p^2}}\right]=-\left(1-
\frac{4u}{\tilde{\omega}_p}+\frac{8u^2}{\tilde{\omega}_p^2}\right)
e^{-\sqrt{u^2+\tilde{\omega}_p^2}}.
\label{eq18}
\end{equation}

Substituting Eq.~(\ref{eq18}) in Eq.~(\ref{eq15}) and performing all integrations
with respect to $u$, we have
\begin{equation}
{\cal F}_{p,\,\rm TE}^{(l=0)}(a,T)=-\frac{k_BT}{16\pi a^2}\left[
\vphantom{\left(\frac{48}{\omega^2}\right)
e^{-\tilde{\omega}_p}}
\tilde{\omega}_p\,e^{-\tilde{\omega}_p} -
4\tilde{\omega}_p\,{K}_2(\tilde{\omega}_p)
+\left(17-
\frac{48}{\tilde{\omega}_p}+\frac{48}{\tilde{\omega}_p^2}\right)
e^{-\tilde{\omega}_p}
\right],
\label{eq19}
\end{equation}
\noindent
where ${K}_n(z)$ is the Bessel function of the imaginary argument.
Using the asymptotic expression for this function at large argument and
preserving only the main term in Eq.~(\ref{eq19}), we obtain
\begin{equation}
{\cal F}_{p,\,\rm TE}^{(l=0)}(a,T)=-\frac{k_BT}{16\pi a^2}
\tilde{\omega}_p
\left(1-\sqrt{\frac{8\pi}{\tilde{\omega}_p}}+\frac{17}{\tilde{\omega}_p}\right)
e^{-\tilde{\omega}_p}.
\label{eq20}
\end{equation}

Note that the first term on the right-hand side of this equation is equal to the main
term of the quantity ${\cal F}_{p,\,\rm TM}^{(l=0)}$
defined in Eq.~(\ref{eq14}). The entire quantity
${\cal F}_{p,\,\rm TE}^{(l=0)}$ decreases exponentially with increasing film
thickness, as does the quantity ${\cal F}_{p,\,\rm TM}^{(l=0)}$.
Notice also that similar expression for the configuration of two metallic plates
separated by a dielectric layer contains the perturbation expansion in powers of
$1/\tilde{\omega}_p$, whereas in Eq.~(\ref{eq20}) the expansion parameter is
$1/\sqrt{\tilde{\omega}_p}$. By summing the main contributions in Eqs.~(\ref{eq14})
and (\ref{eq20}), we find the main term in the Casimir free energy
${\cal F}_{p}^{(l=0)}$ calculated using the plasma model
\begin{equation}
{\cal F}_{p}^{(l=0)}(a,T)=-\frac{k_BT}{8\pi a^2}
\tilde{\omega}_p
e^{-\tilde{\omega}_p}.
\label{eq21}
\end{equation}

We are coming now to the contribution of Matsubara terms with $l\geq 1$ to the
Casimir free energy. It turns out that their role is radically different when the
metallic film is described either by the Drude or by the plasma model.
We start from the case of the Drude model (\ref{eq6}). In this case simple estimations
show (see the Appendix) that for film thicknesses exceeding 110\,nm
the contribution of all Matsubara terms with $l\geq 1$ is negligibly small,
as compared to the zero-frequency term (\ref{eq10}). Thus, for our configuration
of a metallic film sandwiched between two dielectric plates the classical limit is
reached for surprisingly thin films if a metal is characterized by the Drude model
(recall that for two metallic plates separated by a dielectric layer the classical
limit starts at separations exceeding about $6\,\mu$m). This fact does not depend
on a material of the dielectric plates.

When the metallic film is described by the plasma model (\ref{eq7}), numerical
computations show that the contribution of all Matsubara terms with  nonzero frequency
to the Casimir free energy does not become smaller than the classical term (\ref{eq21})
up to very large film thicknesses. Thus, for a Au film of more than $1\,\mu$m
thickness in vacuum at room temperature ($T=300\,$K) the main term in the contribution
of all Matsubara frequencies with $l\geq 1$ is given by (see the Appendix)
\begin{equation}
{\cal F}_{p}^{(l\geq 1)}(a,T)=-\frac{k_BT}{4\pi a^2}
\tilde{\omega}_p e^{-\tilde{\omega}_p}
\sum_{l=1}^{\infty}e^{-\zeta_l^2/(2\tilde{\omega}_p)}.
\label{eq22}
\end{equation}

{}From Eqs.~(\ref{eq21}) and (\ref{eq22}) one finds that for film thicknesses of
6, 30, and $50\,\mu$m the ratio ${\cal F}_{p}^{(l\geq 1)}/{\cal F}_{p}^{(l=0)}$
is equal to 4.95, 1.66, and 1.06, respectively. Only for a film thickness
$a=100\,\mu$m this ratio becomes less than unity (it is equal to 0.46).
But even in this case the classical limit is not yet achieved. Note also that at
$100\,\mu$m the exponential factor $\exp(-\tilde{\omega}_p)$ is equal to
$2.5\times 10^{-3566}$, i.e., the problem has no physical meaning.

It is important to underline that the main contribution to the Casimir free energy
is of quantum origin because the power of the exponent in Eq.~(\ref{eq22})
depends on the Planck constant
\begin{equation}
\frac{\zeta_l^2}{2\tilde{\omega}_p}=
\frac{a(2\pi k_BTl)^2}{c\omega_p\hbar^2}.
\label{eq23}
\end{equation}
\noindent
Thus, in the configuration of a metallic film sandwiched between two dielectric plates
the Casimir free energy has no classical limit if the metal of a film is described
by the plasma model. Such a radical difference with the case of the Drude metallic
film, where the classical limit is reached, but for surprisingly small film
thicknesses, deserves a discussion.

To address this point, we consider the behavior of the Casimir free energy in the
limiting case $\omega_p\to\infty$. In this limit the metal of a film is usually
supposed to turn into an ideal metal characterized by the infinitely large magnitude
of dielectric permittivity at all frequencies. At the surface of an ideal metal the
tangential component of electric field, as well as the normal component of magnetic
induction, must vanish. This reflects the fact that electromagnetic oscillations cannot
penetrate in the interior of an ideal metal film and, thus, the Casimir free energy
of such a film must be equal to zero.

It is interesting to verify whether the above results, obtained using the Drude and
the plasma models, satisfy this physical requirement. In the case of the plasma model
from Eqs.~(\ref{eq21}) and (\ref{eq22}) we immediately arrive at
\begin{equation}
\lim_{\omega_p\to\infty}{\cal F}_{p}(a,T)=0.
\label{eq24}
\end{equation}
\noindent
Thus, if real metal is described by the plasma model, the Casimir free energy
caused by quantum fluctuations of the electromagnetic field, vanishes when the metal
becomes an ideal one, as it should be.

Another situation holds for the Drude model. In this case
from Eqs.~(\ref{eq10}) and (\ref{A8}) we find that
\begin{equation}
\lim_{\omega_p\to\infty}{\cal F}_{D}(a,T)=-\frac{k_BT}{16\pi a^2}\zeta(3),
\label{eq25}
\end{equation}
\noindent
i.e., the ideal metal film is characterized by a nonzero Casimir free energy
contrary to physical intuition.

In the next sections the above results are supported by numerical computations performed
for real metal films both in vacuum and situated between two dielectric plates made of some
specific dielectric material. It should be pointed out that similar results also hold
for the configuration of a metallic film deposited on a dielectric substrate.

\section{Computations of the Casimir free energy of gold films}

Here, we compute the Casimir free energy of a Au film. The dielectric permittivity of Au
$\varepsilon_l^{(0)}$ at the imaginary Matsubara frequencies is obtained from the tabulated
optical data for the complex index of refraction \cite{57} extrapolated to zero frequency using
either the Drude or the plasma model (see Refs.~\cite{1,3} for details).
The film is either in vacuum or sandwiched between two thick sapphire plates (thicker than
about $2\,\mu$m plate can be already considered as a semispace with respect to the Casimir
effect \cite{3}). For the dielectric permittivity of sapphire at the imaginary Matsubara
frequencies there is rather precise analytic expression \cite{58}
\begin{equation}
\varepsilon_l^{(\pm 1)}=1+\frac{C_{\rm IR}\,\omega_{\rm IR}^2}{\omega_{\rm IR}^2+\xi_l^2}
+\frac{C_{\rm UV}\,\omega_{\rm UV}^2}{\omega_{\rm UV}^2+\xi_l^2},
\label{eq26}
\end{equation}
\noindent
where
$C_{\rm UV}=2.072$, $C_{\rm IR}=7.03$, $\omega_{\rm UV}=2.0\times 10^{16}\,$rad/s,
and $\omega_{\rm IR}=1.0\times 10^{14}\,$rad/s.
All computations have been performed using the Lifshitz formula (\ref{eq1}) written in terms
of dimensionless variables $u$ and $\zeta_l$ [see Eq.~(\ref{A1})].

In Fig.~1 we plot
the  magnitudes of the Casimir free energy per unit area at $T=300\,$K
computed using the Drude (the dashed lines) and the plasma (the solid lines)
model approaches as the functions of film thickness in the configurations
 of a Au film sandwiched between two sapphire plates (the pair of solid and
 dashed lines numbered 1)
and for a Au film in vacuum  (the pair of lines numbered 2).
The computational results are presented in the double logarithmic scale.

As is seen in Fig.~1, in each configuration the theoretical predictions of the plasma
and Drude model approaches are almost coinciding for the smallest film thickness. With
increasing thickness of a Au film the predictions of both approaches differ considerably.
Thus, for the configuration of a sandwiched film of thicknesses 50, 100, and 200\,nm the
magnitudes of the free energies predicted by the Drude model approach are larger than those
predicted by the plasma model approach by the factors of 1.97, 61.9, and $3.6\times 10^5$,
respectively. In doing so, the values of the Casimir free energy for the films of 50 and
100\,nm thickness calculated according to the plasma model approach are equal to --42.95 and
$-0.1633\,\mbox{pJ/m}^2$, respectively.

For Au films of 50, 100, and 200\,nm thickness in vacuum  the predictions of the Drude model
approach are larger than for the plasma one by the factors of 1.72, 50.45, and $3.17\times 10^5$,
respectively. In this case the respective values of the Casimir free energy for the films of 50 and
100\,nm thickness computed using the plasma model approach are  --58.60 and
$-0.2012\,\mbox{pJ/m}^2$.

It is interesting that the predictions of the Drude model approach for the Casimir free energy
in the two configurations under consideration (the sandwiched Au film and the Au film in vacuum)
are almost coinciding for film thicknesses $a\geq 100\,$nm (they are equal up to 0.4\% at 100\,nm
and have at least four significant figures common at $a\geq 180\,$nm). If, however, the plasma
model approach is used, the magnitudes of the Casimir free energy predicted for the Au films
of thicknesses 100 and 200\,nm in vacuum are larger than those for a sandwiched film by the
factors of 1.23 and 1.15, respectively. This means that when the plasma
model approach is used in computations, the difference in theoretical predictions for different
configurations is preserved for much larger film thicknesses than for the Drude model approach.
This is explained by the fact that for metallic films described by the Drude model the classical
limit is already achieved for the film thickness of about 110\,nm (see Sec.~II).

Now we consider the smaller film thicknesses, where the Casimir free energy does not depend on
whether the plasma or the Drude model approach is used in computations, and investigate the role
of relativistic effects. In the nonrelativistic limit $r_{\rm TE}^{(0,\pm 1)}=0$ and in terms
of the dimensionless variables one obtains
\begin{equation}
{\cal F}_{D(p)}(a,T)=
\frac{k_BT}{8\pi a^2}\,\sum_{l=0}^{\infty}{\vphantom{\sum}}^{\prime}
\int_{0}^{\infty}\!\!\!u\,du
\ln\left[1-r_{\rm TM}^{(0,+1)}(i\zeta_l)
r_{\rm TM}^{(0,-1)}(i\zeta_l)
e^{-u}\right],
\label{eq27}
\end{equation}
\noindent
where
\begin{equation}
r_{\rm TM}^{(0,\pm 1)}(i\zeta_l)=\frac{\varepsilon_l^{(\pm 1)}-
\varepsilon_l^{(0)}}{\varepsilon_l^{(\pm 1)}+\varepsilon_l^{(0)}}.
\label{eq28}
\end{equation}
\noindent
Note that the same computational results for small film thicknesses are obtained if we change
the discrete Matsubara frequencies $\zeta_l$ in Eqs.~(\ref{eq27}) and (\ref{eq28}) with the
continuous variable and make a replacement
\begin{equation}
{k_BT}\sum_{l=0}^{\infty}{\vphantom{\sum}}^{\prime}\to
\frac{\hbar c}{4\pi a}\int_{0}^{\infty}\!\!\!d\zeta.
\label{eq29}
\end{equation}
\noindent
As a result, the quantity (\ref{eq27}) does not depend on $T$ and gives the Casimir energy
per unit area.

In Fig.~2 we present the common computational results, obtained using the Drude or the plasma
model approach for the magnitude of the nonrelativistic Casimir energy (\ref{eq27}) of a Au
film in vacuum, multiplied by the separation squared, as a function of $a$ (the dashed line).
As is seen from Eqs.~(\ref{eq27}) and (\ref{eq28}), in the case
$\varepsilon_l^{(+1)}=\varepsilon_l^{(-1)}$ the free energy ${\cal F}$ is unchanged if one
replaces $\varepsilon_l^{(+1)}$ with $\varepsilon_l^{(0)}$ and vice versa.
This means that in the nonrelativistic limit the dashed line in Fig.~2 also presents the
Casimir energy in the configuration of two Au plates separated with a vacuum gap.

For comparison purposes, the solid lines 1 and 2 in Fig.~2 show the fully relativistic
computational results for the magnitudes of the Casimir free energy per unit area multiplied
by $a^2$ in the configurations of a Au film in vacuum and of two Au plates separated with a vacuum
gap, respectively. As is seen in Fig.~2, even for a smallest film thickness $a=1\,$nm there is
a detectable difference (of approximately 3\,meV) between the nonrelativistic and relativistic
computational results. This difference increases with increasing film thickness.
For example, for a film of 10\,nm thickness the relativistic result for a Au film
in vacuum (the line 1)
differs from the nonrelativistic one by 45\,meV. The relativistic (the line 2) and the
nonrelativistic (the dashed line) results in the configuration
of two Au plates separated with a vacuum gap of 10\,nm width differ by 30\,meV.
Thus, from Fig.~2 it is seen that the relativistic effects play an important role even for the
thinnest Au films.

{}From this figure it is also seen that in the relativistic case the magnitudes of the free
energy of two Au plates separated with a vacuum gap (the line 2) are somewhat larger than for
a Au film in vacuum (the line 1). The difference between them increases with increasing width
of the gap (film thickness) and achieves 15\,meV for $a=10\,$nm.

For the configuration of a Au film sandwiched between two sapphire plates the computational
results for the Casimir free energy and energy are similar to those shown in Fig.~2.

Now we consider the dependence of the Casimir free energy of a metallic film of the fixed
thickness on the temperature. When the plasma model approach is used in computations,
the temperature dependence may
arise only through the summation over the Matsubara frequencies and the factor $k_BT$ in
Eq.~(\ref{eq1}). For the Drude model approach there is also an additional temperature dependence
through the relaxation parameter in Eq.~(\ref{eq6}). Thus, to find the free energy at different
temperatures, one should use this dependence in computations.

As was mentioned above, for Au at $T=300\,$K the relaxation parameter $\gamma(T)=0.035\,$eV.
Within the temperature interval from room temperature down to $T_D/4$, where $T_D$ is the
Debye temperature, the linear dependence $\gamma(T)\sim T$ is preserved (note that for Au
$T_D=165\,$K \cite{59}). In the interval from $T_D/4$ down to the temperature of helium
liquefaction $\gamma(T)\sim T^5$ according to the Bloch-Gr\"{u}neisen law \cite{60}. and
at liquid helium temperatures $\gamma(T)\sim T^2$ holds \cite{59}.

The computations of the Casimir free energy as a function of temperature are performed for
the film thickness $a=55\,$nm, where the predictions of the plasma and Drude model approaches
differ, at room temperature, by roughly a factor of two. The same equations as above have been
used, but the temperature dependence of the relaxation parameter was taken into account.
The computational results are shown in Fig.~3. The pairs of lines numbered 1 and 2 are obtained
for a Au film sandwiched between two sapphire plates and for a Au film in vacuum, respectively.
For both configurations the solid lines are computed using the plasma model approach and the
dashed lines by the Drude model approach.

As is seen in Fig.~3, the magnitudes of the Casimir free energy of a Au film in vacuum are
larger than for a sandwiched film. This is in agreement with Fig.~1. It is seen also that in
the temperature interval from 0\,K to 300\,K the Casimir free energy computed using the plasma
model approach does not depend on temperature for both configurations under consideration.
In fact for all reasonable film thicknesses, when the Casimir free energy is not too small,
the computational results using the plasma model approach do not depend on the temperature.
The dashed lines in Fig.~3 demonstrate the strong linear dependence of the Casimir free energy
on temperature when the Drude model approach is used in computations. This dependence is caused by
the contribution of the zero-temperature term (\ref{eq10}) to the Casimir free energy and by the
fact that the contribution of all Matsubara terms with $l\geq 1$ is rather small even for rather
thin metallic films.

To illustrate the role of temperature in the Drude model approach, in Fig.~4 we plot the
magnitudes of the Casimir free energy versus film thickness computed at $T=300\,$K
(the solid line) and at $T=77\,$K (the dashed line). In agreement with Fig.~3, at larger
temperature the magnitude of the Casimir free energy is larger for each fixed film thickness.
However, for thin films ($a<30\,$nm) the effect of temperature is negligibly small.
For films of thicknesses equal to  50, 100, and 200\,nm the ratio of the Casimir free energy
 at $T=300\,$K to that  at $T=77\,$K is equal to 1.4, 3.7, and 3.9, respectively.
 This demonstrates that the thermal effect predicted by the Drude model approach contributes
 a lot even for not too thick metallic films (note that for $a=200\,$nm the ratio of free
 energies indicated above is equal to the ratio of temperatures $300/77\approx 3.9$).

\section{Casimir pressure for metallic films}

The Casimir pressure is obtained from the Casimir free energy per unit area by the negative
differentiation with respect to the film thickness
\begin{equation}
P(a,T)=-\frac{\partial{\cal F}(a,T)}{\partial a},
\label{eq30}
\end{equation}
\noindent
where ${\cal F}(a,T)$ is given by Eq.~(\ref{eq1}). Using the dimensionless variables, one
obtains
\begin{eqnarray}
&&
P_{D(p)}(a,T)=-\frac{k_BT}{8\pi a^3}\,\sum_{l=0}^{\infty}{\vphantom{\sum}}^{\prime}
\int_{0}^{\infty}\!\!\sqrt{u^2+\varepsilon_{l,D(p)}^{(0)}\zeta_l^2}u\,du
\left\{\left[
\frac{e^{\sqrt{u^2+\varepsilon_{l,D(p)}^{(0)}\zeta_l^2}}}{r_{\rm TM}^{(0,+1)}(i\zeta_l,u)
r_{\rm TM}^{(0,-1)}(i\zeta_l,u)}-1
\right]^{-1}\right.
\nonumber \\[1mm]
&&~~~~~~~~~~
+\left.\left[
\frac{e^{\sqrt{u^2+\varepsilon_{l,D(p)}^{(0)}\zeta_l^2}}}{r_{\rm TE}^{(0,+1)}(i\zeta_l,u)
r_{\rm TE}^{(0,-1)}(i\zeta_l,u)}-1
\right]^{-1}\right\}.
\label{eq31} 
\end{eqnarray}

Similar to the case of the free energy, for the Drude model approach the zero-frequency term in the
Casimir pressure (\ref{eq31}) becomes dominant for film thicknesses exceeding 110\,nm.
Under this condition the Casimir pressure in both configurations of a metallic film
sandwiched between two dielectric plates or for a metallic film in vacuum is given by
\begin{equation}
P_D(a,T)\approx P_{D,\,{\rm TM}}^{(l=0)}(a,T)=-\frac{k_BT}{8\pi a^3}\zeta(3).
\label{eq32}
\end{equation}

For the plasma model approach under the condition $\tilde{\omega}_p\gg 1$ the contribution
of the zero-frequency term to the Casimir pressure is given by
\begin{equation}
P_p^{(l=0)}(a,T)=-\frac{k_BT}{8\pi a^3}
\tilde{\omega}_p^2\,e^{-\tilde{\omega}_p}.
\label{eq33}
\end{equation}
\noindent
In this case, however, the contribution of nonzero Matsubara terms is not small, as compared
to the quantity (\ref{eq33}), for all physically reasonable film thicknesses.
Thus, if the plasma model approach is used in computations, the classical limit for the
Casimir pressure is not reached. In the limiting case $\omega_p\to\infty$ the Casimir
pressure calculated using the plasma model approach vanishes. As to the Drude model approach,
the Casimir pressure for a metallic film goes to the classical limit (\ref{eq32}) for any
film thickness, i.e., is nonzero when $\omega_p\to\infty$. This is in contradiction to the
fact that electromagnetic oscillations cannot penetrate into an ideal metal.

Now we present the results of numerical computations of the Casimir pressure for Au films.
The magnitudes of the Casimir pressure at $T=300\,$K computed using Eq.~(\ref{eq31})
in the framework of the Drude model (the dashed lines) and the plasma model (the solid lines)
approaches as the functions of film thickness are shown in Fig.~5 for a Au film between
two sapphire plates (the pair of lines 1) and for a Au film in vacuum (the pair of lines 2).
Similar to the free energy, for small film thicknesses below 30\,nm the plasma and Drude
model approaches lead to almost coinciding results in both configurations.
However, for the configuration of a Au film between sapphire plates
for film thicknesses of 50, 100, and 200\,nm the magnitudes of the Casimir pressure
calculated using the Drude model approach are larger than those calculated by the plasma
model approach by the factors of 1.33, 12.7, and $3.73\times 10^4$, respectively.
The values of the Casimir pressure for the films of 50 and 100\,nm thickness, calculated by the
plasma model approach, are equal to 5.21 and 0.0172\,Pa, respectively.

According to Fig.~5, the Casimir pressures for Au films of 50, 100, and 200\,nm
thickness in vacuum calculated using the Drude model approach are larger than
those calculated by the plasma
model one by the factors of 1.24, 10.4, and $3.22\times 10^4$, respectively.
For Au films of 50 and 100\,nm thickness the respective values of the Casimir pressure
calculated in the framework of the plasma model approach are equal to 7.318 and 0.0245\,Pa.

Similar to the Casimir free energy, the values of the Casimir pressure predicted by the Drude
model approach for $a\geq 100\,$nm do not depend on the configuration (either a sandwiched Au
film or a Au film in vacuum). For the plasma model approach, however,
some differences in the values
of the Casimir pressure in different configurations are preserved up to much larger film
thicknesses (see Fig.~5). In the same way as for the Casimir free energy, the relativistic effects
contribute considerably to the Casimir pressure even for Au films of smallest thicknesses.

We are coming now to computation of the Casimir pressure at different temperatures.
This is done using Eq.~(\ref{eq31}) and taking into account the dependence of the relaxation
parameter on $T$ (see Sec.~III). The computational results for $a=55\,$nm are shown in Fig.~6,
where the Casimir pressures for a Au film between two sapphire plates and for a Au film in
vacuum are shown versus temperature by the pairs of lines 1 and 2, respectively (the solid lines
are computed using the plasma model approach and the dashed lines by the Drude model approach).

{}From Fig.~6 it is seen that in the temperature region from 0\,K to 300\,K the Casimir
pressures computed using the plasma model approach do not depend on $T$. This is qualitatively
the same as it holds for the Casimir free energy (see Fig.~3) and is explained by the same
reasons. The linear dependence of the Casimir pressure computed using the Drude model approach
on $T$ is caused by the dominant contribution of the zero-frequency term (\ref{eq33}).
Similar to the Casimir free energy, at each fixed temperature the magnitudes of the Casimir
pressure for a Au film in vacuum are somewhat larger than for a Au film sandwiched between
two sapphire plates.

\section{Conclusions and discussion}

In the foregoing, we have investigated the Casimir effect for metallic films between dielectric
plates and found that it possesses unusual properties which have not been discussed in the previous
literature. The most striking feature of this configuration is that the theoretical predictions
of the Lifshitz theory using the Drude and the plasma model approaches differ significantly
even for very thin films of a few tens nanometer thickness. This is different from previous
understanding for the case of two plates separated
with a vacuum gap or a dielectric layer, where a difference
by a factor of two between these approaches appeared only at large separations above
$6\,\mu$m. It is important that in our case the relatively large difference between the
predictions of the two approaches is obtained not for the force differences \cite{52,53,54} (the latter
requires difference force measurements for the experimental test of the predicted effects), but
for the separate values of the free energy and pressure, as this holds at large separation
distances.

Another interesting feature is the conceptual difference between the predictions of the Drude
and plasma model approaches for metallic films. For the previously considered configurations
at large separations (high temperatures) the classical limits were achieved in the framework
of both theoretical approaches, but the values of the Casimir free energy and pressure differed
by a factor of two. We have shown that for metallic films the classical limit is achieved
for thin films of about 100\,nm thickness and only if the Drude model approach is used.
If the metallic film is described by the plasma model approach, the classical limit is
not achieved for any reasonable film thickness and the Casimir effect has an entirely quantum
character. In this case the Casimir free energy and pressure are decreasing exponentially to zero
with increasing film thickness and differences with theoretical predictions of the Drude
model approach quickly reach several orders of magnitude.

The described difference between the theoretical predictions of both approaches thus has the
far-reaching consequences. According to our results, the Casimir free energy of a metallic film
described by the Drude model approach goes to a nonzero classical value when the plasma
frequency goes to infinity. This is contrary to physical intuition which suggests that
electromagnetic fluctuations cannot penetrate into an ideal metal film, and, thus, its Casimir
free energy should be equal to zero. It should be emphasized, that the Casimir free energy
of a metallic film calculated using the plasma model approach vanishes when the plasma
frequency goes to infinity in accordance with physical intuition.

This raises the question as to why the Drude model which
has been successfully tested in thousands of experiments and in
infinitely many technological applications fails to describe the experimental data of the most
precise experiments on measuring the Casimir force, leads to violation of the Nernst heat
theorem in the Lifshitz theory of dispersion forces and to counter intuitive results in the
limiting case of ideal metals. Taking into account that the Drude model is valid for
classical electromagnetic fields, one can guess that the reason is fundamentally related
to the quantum nature of electromagnetic fluctuations giving rise to dispersion forces
(we stress, that even in the classical limit, which holds in the configuration of two plates
separated with a gap, there are quantum corrections to the classical values of the Casimir
free energy and pressure depending on the Planck constant). If this is the case, the apparent
contradiction of the Bohr-van~Leeuwen theorem with the plasma model approach \cite{42}
(see Sec.~I) would simply mean that the electromagnetic fluctuations responsible for
dispersion forces cannot be considered as classical. Then, the Bohr-van~Leeuwen
theorem will play the same role, as it had originally played in the beginning of the last
century by stating that the ferromagnetic properties of metals cannot be explained in the
framework of classical physics.

Finally, it is pertinent to note that the effects described above can be observed not only
in the plane parallel configurations (for solid films it is not difficult to achieve
parallelity), but in the more popular sphere-plate geometry as well. For this purpose,
the solid Au film should be replaced with a liquid metal like  mercury or, more
conveniently, with an alloy of gallium and indium which is liquid at room temperature.
Then, in accordance with the proximity force approximation, the Casimir force is proportional
to the Casimir free energy calculated in Secs.~II and III. This opens opportunities
for further experimental tests of the Lifshitz theory of dispersion forces in novel
configurations and for the resolution of existing problems.

\appendix
\section{}
In this Appendix we determine the contribution of the Matsubara terms with
$l\geq 1$ to the Casimir free energy (\ref{eq1}) when the film material is described
either by the Drude or by the plasma model.

We begin with the Drude model (\ref{eq6}). It is convenient to use the dimensionless
variable $u=2ak_{\bot}$, introduced in Sec.~II, and the dimensionless Matsubara
frequencies $\zeta_l=2a\xi_l/c$. Then the contribution of all Matsubara terms with
$l\geq 1$ to Eq.~(\ref{eq1}) for any dielectric function can be written in the form
\begin{eqnarray}
&&
{\cal F}_{D(p)}^{(l\geq 1)}(a,T)\equiv \frac{k_BT}{8\pi a^2}\,S_{D(p)}(a,T)=
\frac{k_BT}{8\pi a^2}\,\sum_{l=1}^{\infty}\int_{0}^{\infty}\!\!\!u\,du
\label{A1} \\
&&~~
\times\left\{\ln\left[1-r_{\rm TM}^{(0,+1)}(i\zeta_l,u)
r_{\rm TM}^{(0,-1)}(i\zeta_l,u)
e^{-\sqrt{u^2+\varepsilon_{l,D(p)}^{(0)}\zeta_l^2}}\right]\right.
\nonumber \\
&&~~~~
+\left.\ln\left[1-r_{\rm TE}^{(0,+1)}(i\zeta_l,u)
r_{\rm TE}^{(0,-1)}(i\zeta_l,u)
e^{-\sqrt{u^2+\varepsilon_{l,D(p)}^{(0)}\zeta_l^2}}\right]\right\},
\nonumber
\end{eqnarray}
\noindent
where the reflection coefficients are given by Eq.~(\ref{eq2}), but $k_l^{(n)}(k_{\bot})$
are replaced with $k_l^{(n)}(u)=\sqrt{u^2+\varepsilon_l^{(n)}\zeta_l^2}$.

Now we consider the quantity $\tilde{S}_{D}(a,T)$, which is defined in the same way
as $S_{D}(a,T)$ in Eq.~(\ref{A1}), but with the product of the reflection coefficients
replaced with unity. It is evident that
\begin{equation}
|S_{D}(a,T)|<|\tilde{S}_{D}(a,T)|=-2\sum_{l=1}^{\infty}\int_{0}^{\infty}
\!\!\!u\,du\ln\left(1-e^{-\sqrt{u^2+\varepsilon_{l,D}^{(0)}\zeta_l^2}}\right).
\label{A2}
\end{equation}
\noindent
Using Eq.~(\ref{eq6}) rewritten in terms of dimensionless variables,
\begin{equation}
\varepsilon_{l,D}^{(0)}=1+
\frac{\tilde{\omega}_p^2}{\zeta_l(\zeta_l+\tilde{\gamma})},
\label{A3}
\end{equation}
\noindent
where $\tilde{\gamma}=2a\gamma/c$ is the dimensionless relaxation parameter,
it is easily seen that for film thicknesses $a>110\,$nm the quantity
$\varepsilon_{l,D}^{(0)}\zeta_l^2$ remains large for all $l\geq 1$.
Expanding the right-hand side of Eq.~(\ref{A2}) in powers of a small exponent
and integrating with respect to $u$, one arrives at
\begin{equation}
|S_{D}(a,T)|<2\sum_{l=1}^{\infty}\sum_{n=1}^{\infty}
\frac{1}{n^3}\left(1+n\sqrt{\varepsilon_{l,D}^{(0)}}\zeta_l\right)
e^{-n\sqrt{\varepsilon_{l,D}^{(0)}}\zeta_l}.
\label{A4}
\end{equation}
\noindent
Note that the account of the contribution of core electrons in the dielectric
permittivity $\varepsilon_{l,D(p)}^{(0)}$ would only increase its value and, thus,
decrease the upper bound of $|S_D|$ in Eq.~(\ref{A4}).

Using Eq.~(\ref{A3}) and an evident equality $\zeta_l=\zeta_1 l$, the quantity
$\sqrt{\varepsilon_{l,D}^{(0)}}\zeta_l$ can be represented as
\begin{equation}
\sqrt{\varepsilon_{l,D}^{(0)}}\zeta_l=
\sqrt{\zeta_l^2+\tilde{\omega}_p^2\frac{l}{l+\frac{\tilde{\gamma}}{\zeta_1}}}.
\label{A5}
\end{equation}
\noindent
It is well known that at room temperature
$\tilde{\gamma}/\zeta_1=\gamma/\xi_1<1$. For instance, for Au we have
$\gamma=0.035\,$eV and $\tilde{\gamma}/\zeta_1=0.21$.
As a result, the minimum value of the ratio in Eq.~(\ref{A5}) is achieved
at $l=1$:
\begin{equation}
\min_{l}\frac{l}{l+\frac{\tilde{\gamma}}{\zeta_1}}\approx
\frac{1}{1+0.21}\approx 0.82.
\label{A6}
\end{equation}

On the other hand, the largest value of the same ratio, achieved at $l\to\infty$,
is equal to unity. Now we reinforce the inequality (\ref{A4}) by using
Eq.~(\ref{A6}) in the power of the exponent, but substituting the largest value of
the ratio in the pre-exponent:
\begin{equation}
|S_{D}(a,T)|<2\sum_{l=1}^{\infty}\sum_{n=1}^{\infty}
\left(\frac{1}{n^3}+\frac{1}{n^2}\sqrt{\zeta_l^2+\tilde{\omega}_p^2}\right)
e^{-n\sqrt{\zeta_l^2+0.82\tilde{\omega}_p^2}}.
\label{A7}
\end{equation}

Calculating the sum with respect to $n$, we obtain
\begin{equation}
|S_{D}(a,T)|<2\sum_{l=1}^{\infty}\left[
{\rm Li}_3\left(e^{-\sqrt{\zeta_l^2+0.82\tilde{\omega}_p^2}}\right)
+\sqrt{\zeta_l^2+\tilde{\omega}_p^2}
{\rm Li}_2\left(e^{-\sqrt{\zeta_l^2+0.82\tilde{\omega}_p^2}}\right)
\right].
\label{A8}
\end{equation}
\noindent
We have computed the right-hand side of Eq.~(\ref{A8}) at $T=300\,$K.
At $a=110\,$nm it is equal to 0.058 and quickly decreases with increase
of separation. Thus, at $a=120$ and 150\,nm
the right-hand side of Eq.~(\ref{A8}) is equal to 0.026 and 0.0024,
respectively, and goes to zero at large separations.
For comparison purposes, the contribution of the zero-frequency term
\begin{equation}
\frac{8\pi a^2}{k_BT}\,|{\cal F}_D^{(l=0)}(a,T)|=\frac{1}{2}\zeta(3)
\approx 0.601,
\label{A9}
\end{equation}
\noindent
as it follows from Eqs.~(\ref{eq8}) and (\ref{eq10}). Thus, for metallic
films described by the Drude model, the classical limit holds for all film
thicknesses exceeding 110\,nm, as is stated in Sec.~II.

Now  we consider the contribution of all Matsubara terms with
$l\geq 1$ to the Casimir free energy calculated using the plasma model.
In analytic calculations we restrict ourselves by the case of a Au film
in vacuum at $T=300\,$K. We assume that the film is sufficiently thick
($a\geq 1\,\mu$m). In this case $\tilde{\omega}_p\geq 91$.
Taking into account that the Matsubara frequencies, giving the main
contribution to the Casimir free energy, satisfy the condition
$\xi_l\lesssim 10c/(2a)$, i.e., $\zeta_l\lesssim 10$, we arrive at the
small parameter
\begin{equation}
\frac{\zeta_l}{\sqrt{\zeta_l^2+\tilde{\omega}_p^2}}\ll 1.
\label{A10}
\end{equation}

In terms of dimensionless variables the plasma model is obtained from
Eq.~(\ref{A3}) by putting $\tilde{\gamma}=0$
\begin{equation}
\varepsilon_{l,p}^{(0)}=1+
\frac{\tilde{\omega}_p^2}{\zeta_l^2}.
\label{A11}
\end{equation}
\noindent
We substitute Eq.~(\ref{A11}) in Eq.~(\ref{eq4}), where $k_{\bot}$ is replaced
with $u/(2a)$, and expanding in powers of the parameter (\ref{A10}) find
\begin{equation}
r_{\rm TM}^{(0.\pm1)}(i\zeta_l,u)\approx -1+
\frac{2}{\sqrt{1+\frac{u^2}{\zeta_l^2}}}\,
\frac{\zeta_l}{\sqrt{\zeta_l^2+\tilde{\omega}_p^2}},
\qquad
r_{\rm TE}^{(0.\pm1)}(i\zeta_l,u)\approx 1-
2{\sqrt{1+\frac{u^2}{\zeta_l^2}}}\,
\frac{\zeta_l}{\sqrt{\zeta_l^2+\tilde{\omega}_p^2}}.
\label{A12}
\end{equation}

Taking into account that under our conditions
$\sqrt{1+u^2/\zeta_l^2}\sim 1$, in the zeroth order of the small parameter (\ref{A10})
we can replace the TM and TE reflection coefficients with --1 and 1, respectively.
As a result, the normalized contribution of all Matsubara terms
with $l\geq 1$ in the Casimir
free energy [see Eq.~(\ref{A1})] is given by
\begin{equation}
S_{p}(a,T)=2\sum_{l=1}^{\infty}\int_{0}^{\infty}
\!\!\!u\,du\ln\left(1-e^{-\sqrt{u^2+\zeta_l^2+\tilde{\omega}_p^2}}\right).
\label{A13}
\end{equation}
\noindent
After the expansion in powers of the small exponent and integrating with respect
to $u$, we obtain
\begin{equation}
S_{p}(a,T)=-2\sum_{l=1}^{\infty}\sum_{n=1}^{\infty}
\frac{1}{n^3}\left(1+n\sqrt{\zeta_l^2+\tilde{\omega}_p^2}\right)
e^{-n\sqrt{\zeta_l^2+\tilde{\omega}_p^2}}.
\label{A14}
\end{equation}
\noindent
The exponents in this equation decrease very fast due to the large values
of $\tilde{\omega}_p>91$. Because of this, we can omit all terms with $n\geq 2$
without loss of precision
\begin{equation}
S_{p}(a,T)=-2\sum_{l=1}^{\infty}
\left(1+\sqrt{\zeta_l^2+\tilde{\omega}_p^2}\right)
e^{-\sqrt{\zeta_l^2+\tilde{\omega}_p^2}}.
\label{A15}
\end{equation}
\noindent
Expanding here in powers of the small parameter $\zeta_l/\tilde{\omega}_p$,
we get
\begin{equation}
S_{p}(a,T)=-2\tilde{\omega}_p\sum_{l=1}^{\infty}
\left(1+\frac{1}{\tilde{\omega}_p}+\frac{\zeta_l^2}{2\tilde{\omega}_p^2}\right)
e^{-\tilde{\omega}_p-\zeta_l^2/(2\tilde{\omega}_p)}
\label{A16}
\end{equation}
\noindent
and neglecting by ${1}/{\tilde{\omega}_p}$ and
$\zeta_l^2/(2\tilde{\omega}_p^2)$ as compared with unity, finally arrive at
\begin{equation}
S_{p}(a,T)=-2\tilde{\omega}_pe^{-\tilde{\omega}_p}\sum_{l=1}^{\infty}
e^{-\zeta_l^2/(2\tilde{\omega}_p)}.
\label{A17}
\end{equation}
\noindent
After multiplication by the normalization factor $k_BT/(8\pi a^2)$,
one obtains from Eq.~(\ref{A17}) the desired Eq.~(\ref{eq22}).


\begin{figure}[b]
\vspace*{-4cm}
\centerline{\hspace*{1cm}
\includegraphics{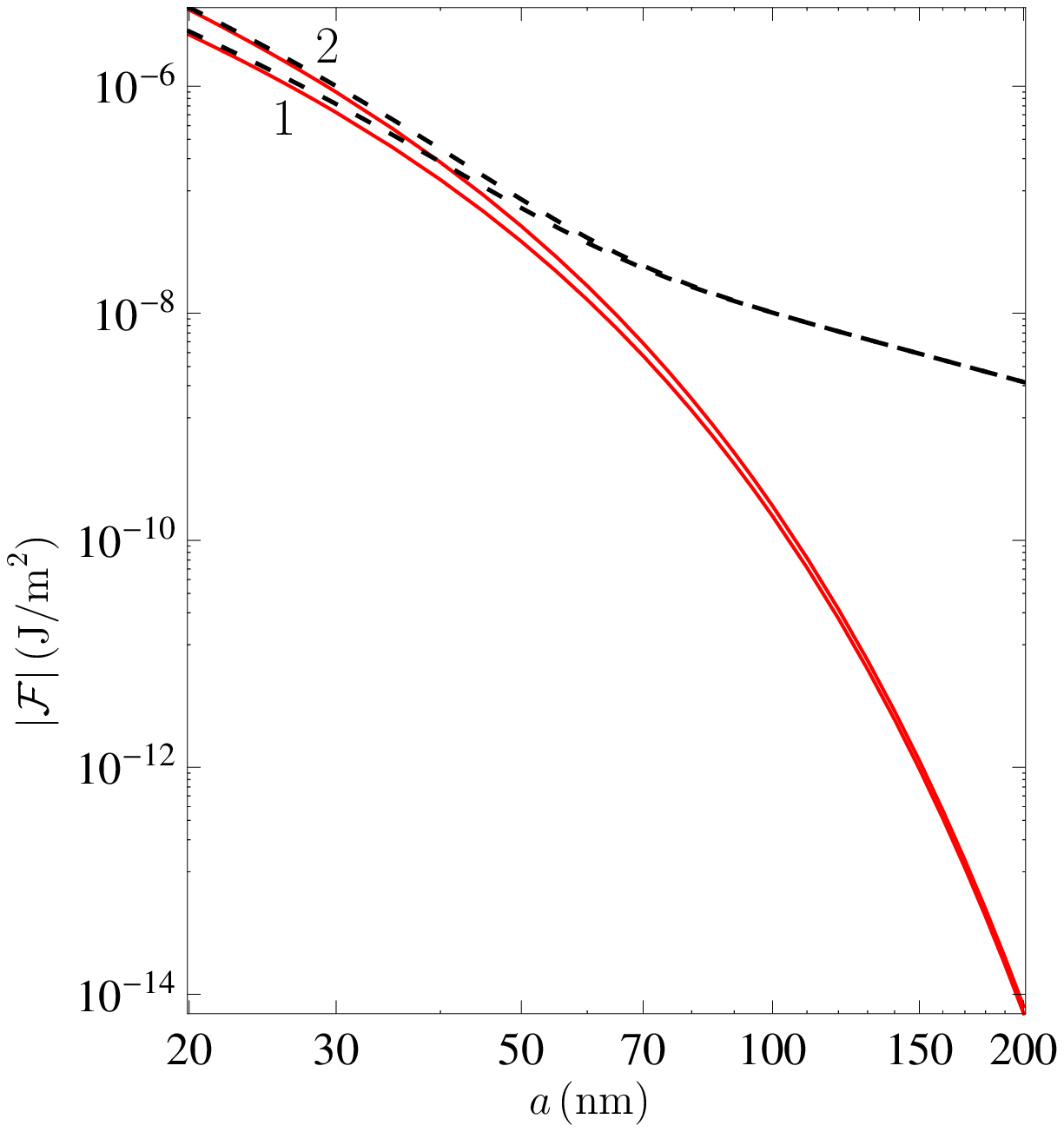}
}
\vspace*{-10cm}
\caption{\label{fg1}(Color online)
The  magnitudes of the Casimir free energy per unit area at $T=300\,$K
computed using the Drude model (the dashed lines) and the plasma model (the solid lines)
 approaches are plotted in the double logarithmic scale versus the film
thickness for a Au film sandwiched between two sapphire plates (the pair of lines 1)
and for a Au film in vacuum  (the pair of lines 2).
}
\end{figure}
\begin{figure}[b]
\vspace*{-8cm}
\centerline{\hspace*{1cm}
\includegraphics{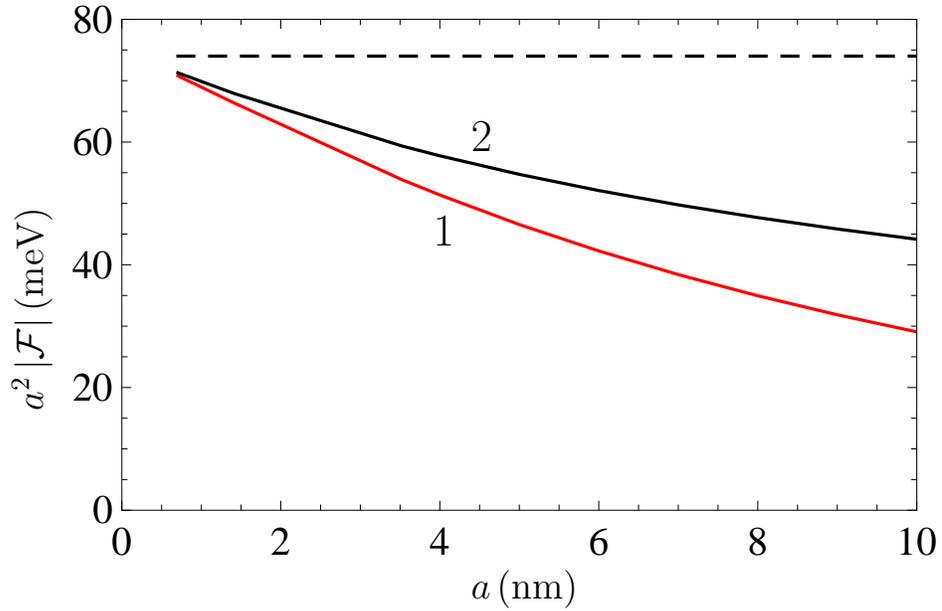}
}
\vspace*{-10cm}
\caption{\label{fg2}(Color online)
The  magnitudes of the Casimir free energy per unit area at $T=300\,$K
are plotted for a Au film in vacuum versus the film thickness (the line 1)
and for two Au plates separated with a vacuum gap versus the gap width
(the line 2). The common nonrelativistic results for the Casimir
 energy per unit area in these configurations are shown by the dashed line.
}
\end{figure}
\begin{figure}[b]
\vspace*{-8cm}
\centerline{\hspace*{1cm}
\includegraphics{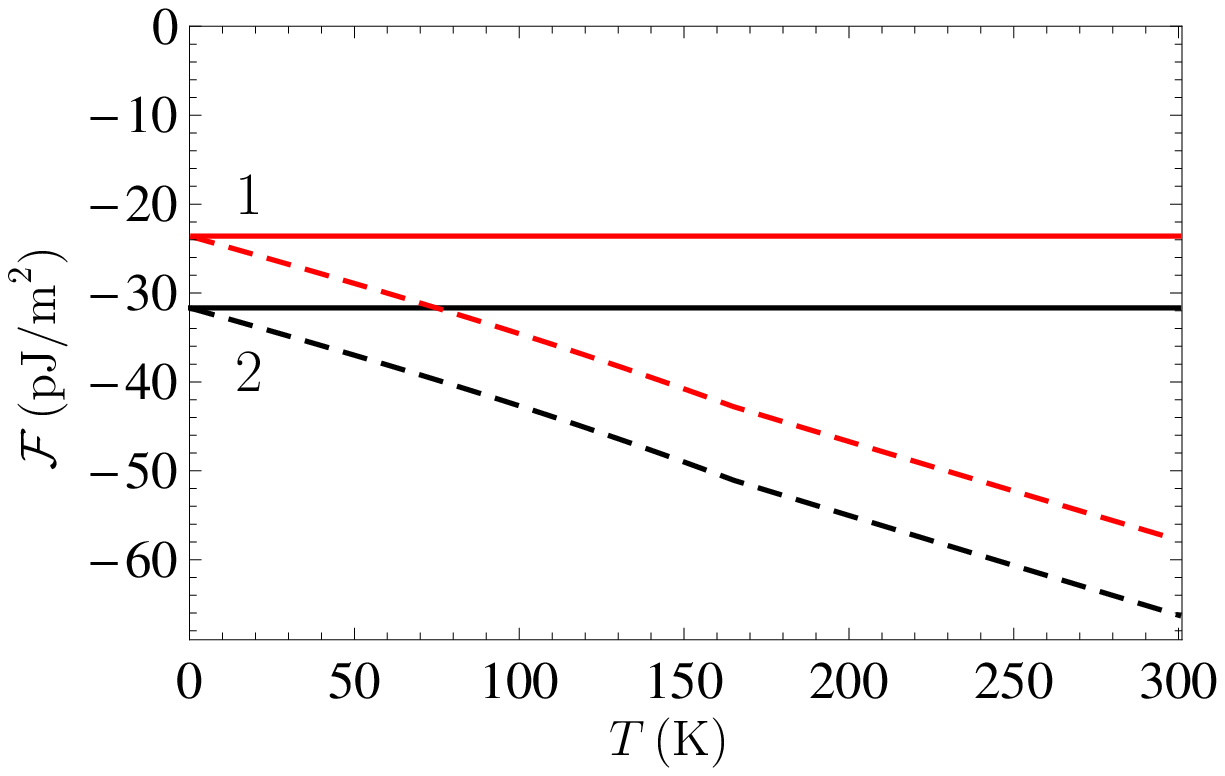}
}
\vspace*{-10cm}
\caption{\label{fg3}(Color online)
The  Casimir free energy per unit area for the film thickness $a=55\,$nm
computed using the Drude model (the dashed lines) and the plasma model (the solid lines)
 approaches are plotted versus temperature
for a Au film sandwiched between two sapphire plates (the pair of lines 1)
and for a Au film in vacuum  (the pair of lines 2).
}
\end{figure}
\begin{figure}[b]
\vspace*{-4cm}
\centerline{\hspace*{1cm}
\includegraphics{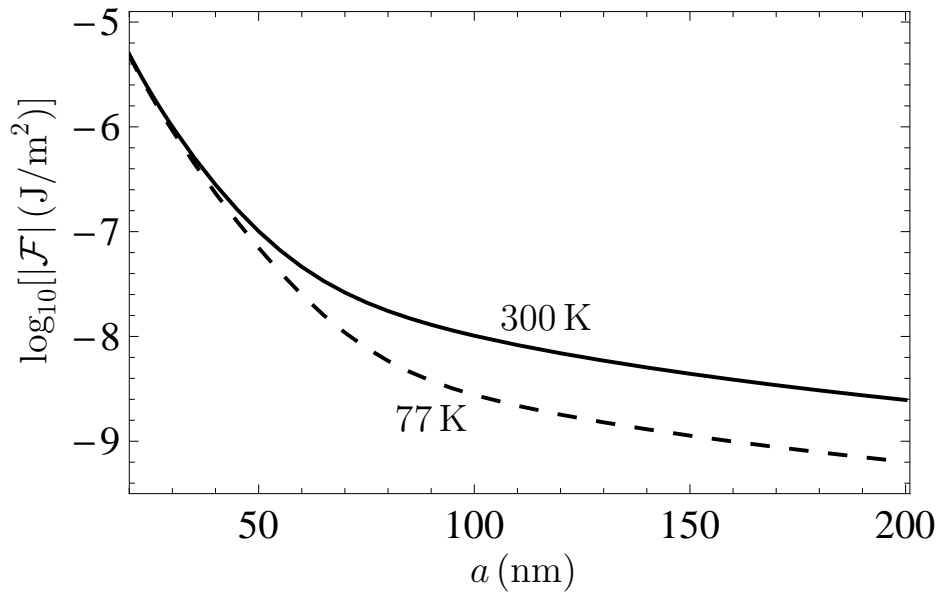}
}
\vspace*{-10cm}
\caption{\label{fg4}
The  magnitudes of the Casimir free energy per unit area at $T=300\,$K
(the solid line) and $T=77\,$K (the dashed lines)
computed using the Drude model approach for a Au film in vacuum
are plotted  versus the film thickness.
}
\end{figure}
\begin{figure}[b]
\vspace*{-4cm}
\centerline{\hspace*{1cm}
\includegraphics{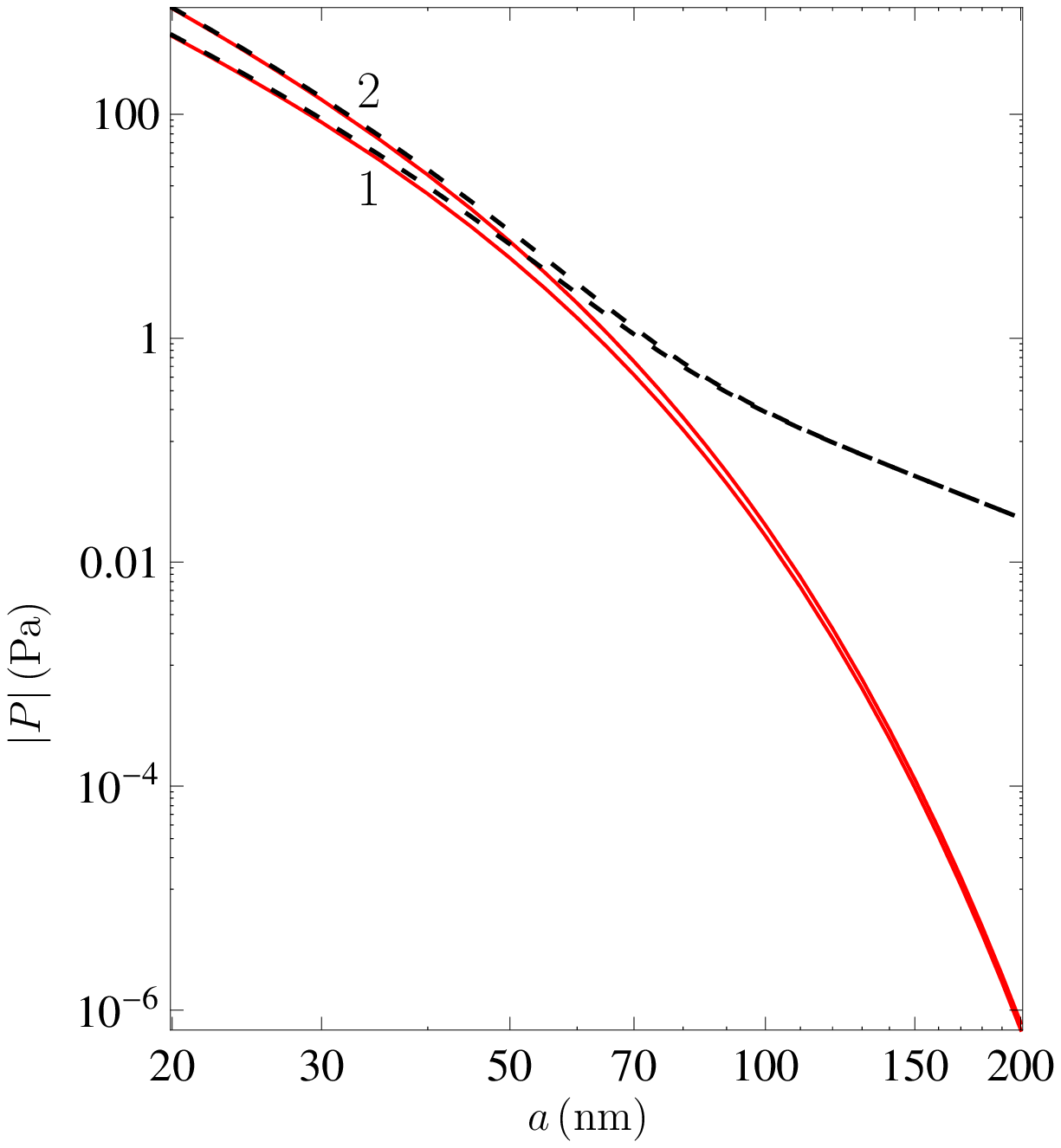}
}
\vspace*{-10cm}
\caption{\label{fg5}(Color online)
The  magnitudes of the Casimir pressure at $T=300\,$K
computed using the Drude model (the dashed lines) and the plasma model (the solid lines)
 approaches are plotted in the double logarithmic scale versus the film
thickness for a Au film sandwiched between two sapphire plates (the pair of lines 1)
and for a Au film in vacuum  (the pair of lines 2).
}
\end{figure}
\begin{figure}[b]
\vspace*{-8cm}
\centerline{\hspace*{1cm}
\includegraphics{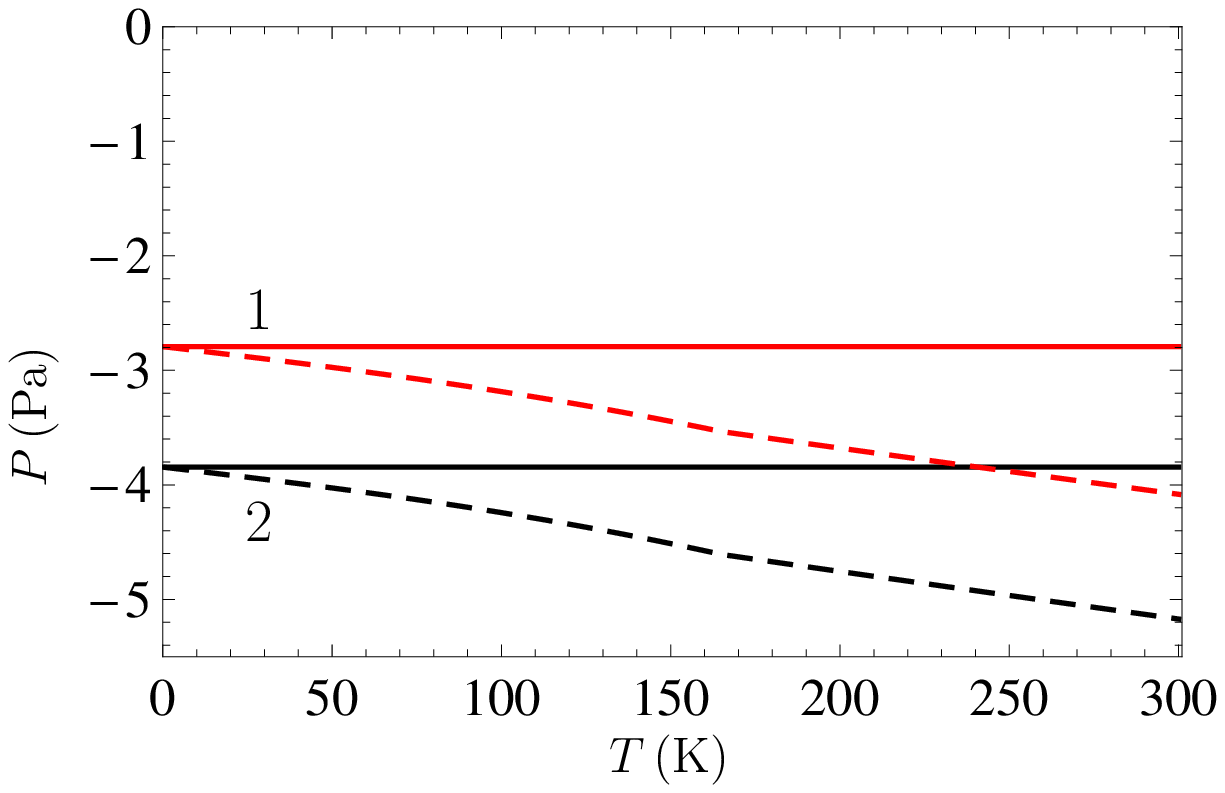}
}
\vspace*{-10cm}
\caption{\label{fg6}(Color online)
The  Casimir pressure for the film thickness $a=55\,$nm
computed using the Drude model (the dashed lines) and the plasma model (the solid lines)
 approaches are plotted versus temperature
for a Au film sandwiched between two sapphire plates (the pair of lines 1)
and for a Au film in vacuum  (the pair of lines 2).
}
\end{figure}
\end{document}